\documentclass[sigconf]{acmart}

\usepackage[normalem]{ulem}
\usepackage{enumitem}

\AtBeginDocument{%
  \providecommand\BibTeX{{%
    \normalfont B\kern-0.5em{\scshape i\kern-0.25em b}\kern-0.8em\TeX}}}

\copyrightyear{2023} 
\acmYear{2023} 
\setcopyright{acmcopyright}\acmConference[WSDM '23]{Proceedings of the Sixteenth ACM International Conference on Web Search and Data Mining}{February 27-March 3, 2023}{Singapore, Singapore}
\acmBooktitle{Proceedings of the Sixteenth ACM International Conference on Web Search and Data Mining (WSDM '23), February 27-March 3, 2023, Singapore, Singapore}
\acmPrice{15.00}
\acmDOI{10.1145/3539597.3573029}
\acmISBN{978-1-4503-9407-9/23/02}

\acmSubmissionID{wsdmsdp16}

\usepackage{booktabs} 
\usepackage{color}

\begin{document}

\title{UserSimCRS: A User Simulation Toolkit for Evaluating  Conversational Recommender Systems} 

\author{Jafar Afzali}
\affiliation{%
  \institution{University of Stavanger}
}
\email{j.afzali@stud.uis.no}

\author{Aleksander Mark Drzewiecki}
\affiliation{%
  \institution{University of Stavanger}
}
\email{am.drzewiecki@stud.uis.no}

\author{Krisztian Balog}
\affiliation{%
  \institution{University of Stavanger}
  \city{Stavanger}
  \country{Norway}
}
\email{krisztian.balog@uis.no}

\author{Shuo Zhang}
\affiliation{%
  \institution{Bloomberg}
  \city{London}
  \country{United Kingdom}
}
\email{szhang611@bloomberg.net}

\renewcommand{\shortauthors}{Jafar Afzali, Aleksander Mark Drzewiecki, Krisztian Balog, \& Shuo Zhang}

\begin{abstract}
We present an extensible user simulation toolkit to facilitate automatic evaluation of conversational recommender systems.  It builds on an established agenda-based approach and extends it with several novel elements, including user satisfaction prediction, persona and context modeling, and conditional natural language generation.  We showcase the toolkit with a pre-existing movie recommender system and demonstrate its ability to simulate dialogues that mimic real conversations, while requiring only a handful of manually annotated dialogues as training data. 
\end{abstract}



\maketitle

\section{Introduction}

Conversational recommender systems (CRSs) elicit user preferences via multi-turn real-time interactions using natural language~\citep{Gao:2021:AIOpen,Jannach:2021:ACMCS}.
There has been a great deal of progress in recent years on various aspects, including question-based user preference elicitation~\citep{Kostric:2021:RecSys,Christakopoulou:2016:KDD,Zou:2020:SIGIR}, multi-turn conversational recommendation strategies~\citep{Lei:2020:WSDM}, and natural language understanding and generation~\citep{Li:2018:NIPS,Zhang:2018:CIKM}.
A major challenges that remains, however, is evaluation~\citep{Gao:2021:AIOpen}.  Due to the dynamic nature of interactions, measuring performance on the conversation level is not possible using offline test collections.  While online evaluation with users of a live service is an option, it is expensive and does not scale.
A promising solution to these issues is \emph{user simulation}~\citep{Balog:2021:DESIRES,Gao:2021:AIOpen}.
The idea there is to build a simulated user that mimics how a real human would respond in a given dialogue situation~\citep{Schatzmann:2006:SSU,Zhang:2020:KDD}. 
Simulation thus offers a repeatable and reproducible means of evaluation.  (We note that it is not meant to replace, but rather to complement human evaluation.)

There is indeed an emerging focus in recent research on using simulation for evaluating conversational information access systems in general~\citep{Balog:2021:DESIRES,Salle:2021:ECIR,Sun:2021:SUS,Sekulic:2022:WSDM,Balog:2021:SIGIRForum} and conversational recommenders in particular~\citep{Zhang:2020:KDD,Zhang:2022:SIGIR}.
The current work aims to contribute to the development of novel CRSs by recognizing the need for better tooling for user simulation.  In particular, we provide an extensible open-source toolkit that is designed specifically for evaluation.
Our work is unique in at least three regards.
First, it focuses on the task of \emph{conversational} recommendation and hence place a strong emphasis on both the recommendation-specific conversation flow and on the human-likeness of the generated user utterances. 
Second, it centers around \emph{evaluation} as opposed to other uses of simulation (most commonly, synthetic data generation for reinforcement learning).
Third, it is designed to work with \emph{existing CRSs}, without needing access to source code or knowledge of their inner workings.  It merely requires collecting and annotating a small sample of dialogues.

\begin{figure*}[!h]
    \centering
    \vspace*{-\baselineskip}
    \includegraphics[width=0.65\textwidth]{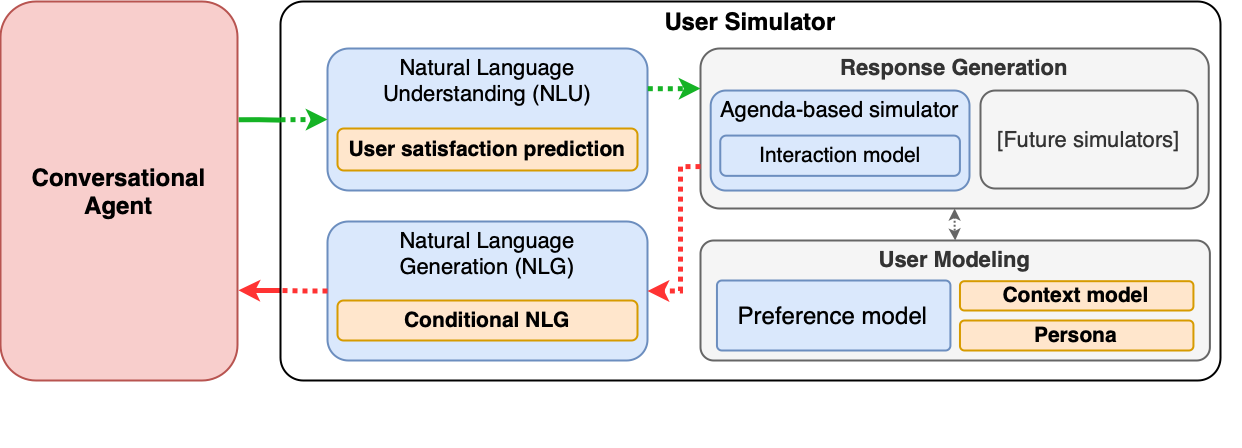}
    \vspace*{-1.5\baselineskip}
    \caption{Conceptual overview of the user simulator. The parts in blue follow~\citep{Zhang:2020:KDD}, while the yellow ones are novel additions.}
    \label{fig:conceptual_overview}
\end{figure*}

Building on an established agenda-based simulator~\citep{Zhang:2020:KDD}, we introduce novel components, motivated by recent research~\citep{Salle:2021:ECIR,Sun:2021:SUS,Zhang:2022:SIGIR}, for modeling user satisfaction, persona and context, and conditional natural language generation.
Given its modular design, the toolkit can also be easily extended with other modeling options or additional components.
The toolkit is comprised of two Python libraries, which are made publicly available on GitHub: DialogueKit\footnote{ \url{https://github.com/iai-group/DialogueKit}} is a collection of generic and reusable dialogue components, and UserSimCRS\footnote{\url{https://github.com/iai-group/UserSimCRS}} is an extensible user simulator built on top.

\section{Related work}
\label{sec:related}

While there are several efforts on simulation toolkits for recommender systems~\citep{Krauth:2020:arXiv,Rohde:2018:arXiv,Shi:2019:RecSys,Ie:2019:arXiv,Mladenov:2021:arXiv}, our work differs from those in two major ways.
First, we focus on the task of \emph{conversational} recommendations and hence place a strong emphasis on natural language understanding and generation.  Thus, unlike others  that operate in the ``intent space,'' we operate in the ``language space.''
Second, our objective is \emph{system evaluation}, as opposed to training end-to-end recommender systems using reinforcement learning (RL).

Our toolkit implements an agenda-based simulator~\citep{Schatzmann:2007:NAACL}, building on and extending the approach in~\citep{Zhang:2020:KDD}. 
Alternatively, model-based simulation could also be employed as it has been done recently for task-based dialogue systems.
\citet{Shi:2019:EMNLP} demonstrate how to build model-based user simulators that rely on a simple Seq2seq dialogue system with copy and attention mechanisms, to facilitate RL-based dialogue system training.
ConvLab-2~\citep{Zhu:2020:ACL} is an open-source toolkit that enables researchers to build task-oriented dialogue systems, where user simulators are provided to support end-to-end evaluation. These simulators can be assembled by equipping a neural network-based user policy with NLU and NLG components.
\citet{Tseng:2021:ACL} propose a learning framework for developing dialogue systems that perform joint optimization with an LSTM-based user simulator, which consists of a dialogue manager, an NLG model, and a dialogue context encoder. The dialogue systems and user simulator models are pre-trained using supervised learning and then fine-tuned using reinforcement learning based on the generated dialogues.
Importantly, such model-based approaches can also be incorporated into our framework in the future.
\section{Conceptual Overview} 
\label{sec:usersim}

The goal of user simulation is to mimic how real users would respond in given dialogue situation~\citep{Schatzmann:2006:SSU,Zhang:2020:KDD}.
Conceptually, our user simulator follows the architecture of a typical task-based dialogue system, which consists of natural language understanding, response generation, and natural language generation components.  Additionally, there is a dedicated user modeling component; see Fig.~\ref{fig:conceptual_overview}.
We opt for a modular design, as opposed to an end-to-end trainable system, in order to have complete control over how responses are generated and to allow for flexible extensions.
Our work builds on and extends the approach proposed in~\citep{Zhang:2020:KDD} as detailed below.

\textbf{\emph{Natural language understanding (NLU)}} is responsible for obtaining a structured representation of text utterances.  Conventionally, it entails intent classification and entity recognition.  Additionally, motivated by recent research~\citep{Salle:2021:ECIR,Sun:2021:SUS}, we also include a classifier for user satisfaction prediction.\footnote{User satisfaction prediction is only used in the training stage to annotate dialogues.}

\textbf{\emph{Response generation}} is currently based on agenda-based simulation~\citep{Schatzmann:2007:NAACL}, however, it could be replaced with other approaches in the future.  Following~\citep{Zhang:2020:KDD}, response generation is based on an \emph{interaction model}, which is responsible for initializing the agenda and updating it. Updates to the agenda can be summarized as follows: if the agent responds in an expected manner, the interaction model \emph{pulls} the next action off the agenda; otherwise, it either repeats the same action as the previous turn or samples a new action.  

\textbf{\emph{User modeling}} consists of three sub-components.  
The \emph{preference model} captures users' likes and dislikes.  Following~\citep{Zhang:2020:KDD}, it is modeled as a \emph{personal knowledge graph}~\citep{Balog:2019:ICTIR}, where nodes correspond to items and attributes.  Novel to our work is the modeling of \emph{persona}, which can capture user-specific traits, e.g., patience or cooperativeness, and \emph{context}, which can characterize the situation of the user, e.g., temporal (time of the day and weekday vs. weekend), relational (alone vs. group setting), or conversational (user satisfaction). 
We focus on contextual aspects as these represent a so far unexplored area of user modeling~\citep{Jannach:2021:ACMCS} and there is evidence suggesting that language usage depends on persona and context~\citep{Sun:2021:SUS,Park:2021:CHB}.

\textbf{\emph{Natural language generation (NLG)}} is currently template-based, that is, given the output of the response generation module, a fitting textual response is chosen and may be instantiated with preferences.  Additionally, we extend the NLG such that it can be conditioned on context.  For example, user responses might be shorter/longer depending on the time of the day or users could use a stronger language when getting dissatisfied with the system.

\section{Software Architecture}
\label{sec:system}

\begin{figure}[!t]
    \centering
    \vspace*{-0.5\baselineskip}
    \includegraphics[width=0.47\textwidth]{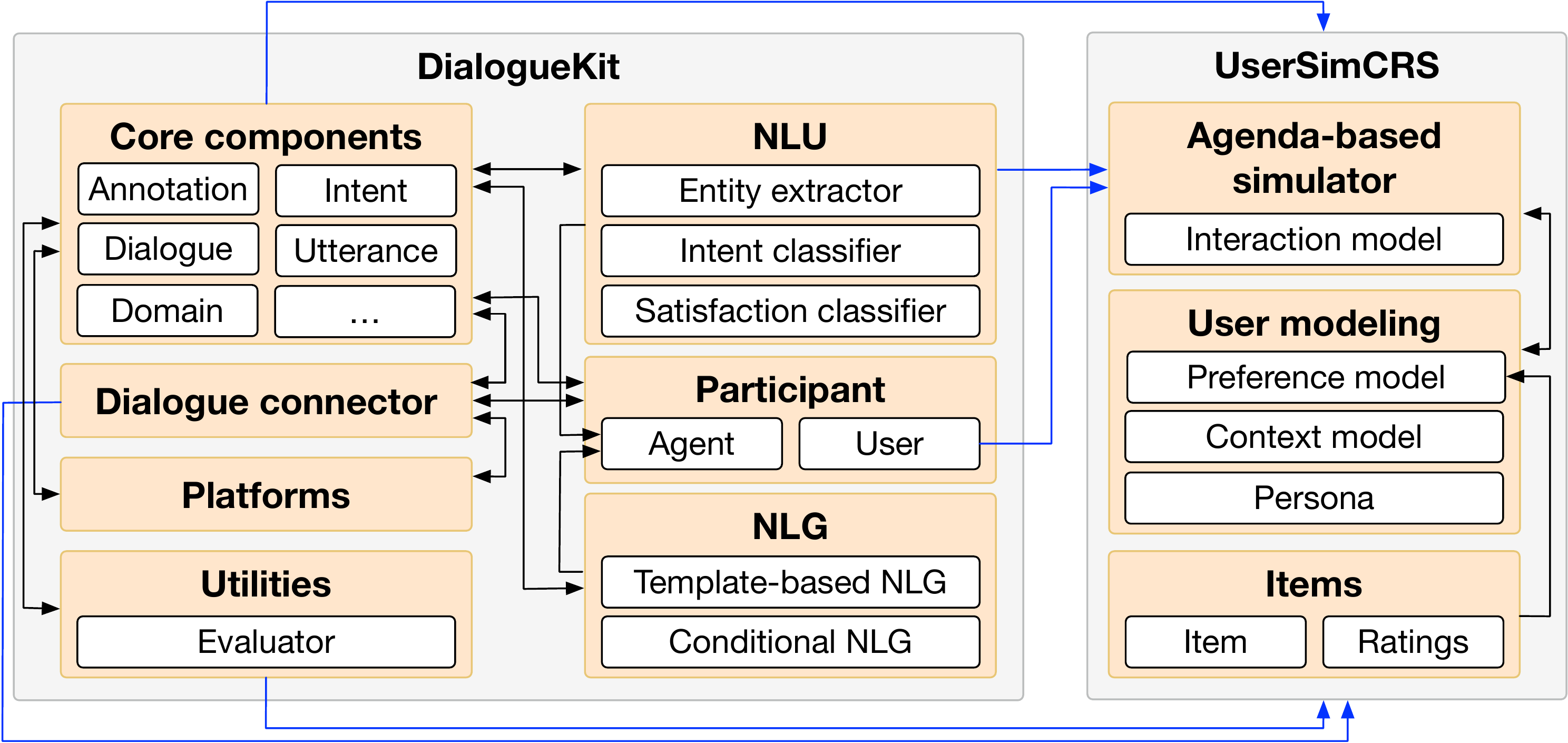}
    \vspace*{-0.5\baselineskip}
    \caption{Overview of the main packages (in yellow) with some of the core modules highlighted (in white).  Arrows indicate intra-library dependencies (in blue) and inter-library dependencies (in black).}
    \label{fig:packages_overview}
    \vspace*{-0.5\baselineskip}
\end{figure}

The toolkit is written in Python and is based on a modular architecture to support additional components, different models, and custom features to be added in the future.
There are two main libraries that are stacked on each other: \textbf{DialogueKit} provides basic dialogue management functionalities, while \textbf{UserSimCRS} contains simulation-specific models and logic. See Fig.~\ref{fig:packages_overview} for an overview of the main packages and their dependencies. 
Both libraries are made available in the Python Package Index (PyPI).

\subsection{DialogueKit}

DialogueKit models \emph{dialogue participants} (users and agents), \emph{domains} (which define the types of slots for a particular application), \emph{utterances}, and \emph{annotations} as base concepts. Utterances may be annotated with \emph{intents} and \emph{slot-value} pairs. DialogueKit currently supports two models for annotation, a cosine classifier for intents and a minimal pipeline DIET classifier~\citep{Tanja:2020:arXiv} for slot-value pairs.\footnote{The DIET classifier can be used for intent detection as well.}
A \emph{dialogue connector} is included to orchestrate and store the conversation between participants (human-human, human-machine, or machine-machine).
Furthermore, the \emph{evaluation} component provides functionality required to evaluate a set of conversations with respect to standard metrics (such as AvgTurns and AvgSuccess).

\subsection{UserSimCRS}

The UserSimCRS library implements the simulation-specific components in Fig.~\ref{fig:conceptual_overview}, specifically, response generation and user modeling. During a conversation, any time the user is asked to provide preferences, the preference model is consulted.  
Context is modeled in a generic way such that it can capture, among others, temporal, relational, and conversational factors. 
The generation of user utterances may be conditioned on the user's context and persona.
Next, we elaborate on how to use UserSimCRS for system evaluation.  Note that the library may also be used for training agents, but that is outside the focus of the current paper.

\section{System Evaluation using Simulation}
\label{sec:eval}

This section discusses how to employ simulation for evaluating an existing CRS and illustrates this with a case study.

\subsection{Methodology}

The main objective of simulation-based evaluation in this work is to establish a \emph{relative comparison} between two systems.  These may be different variants of the same CRS or two different systems.  Importantly, the user simulator needs to target the differences that we care about.  For the sake of illustration, assume that there is a baseline conversational movie recommender that understands movie genres and an improved version that also recognizes plot keywords.  Having a user simulator that asks only for genres but not for plot keywords will not capture the differences between these two systems.  Therefore, as a general principle, the user simulator needs to be co-developed with the CRS and customized to mimic the targeted user behavior. 

\subsection{Setting up Simulation}

A unique feature of our toolkit is that it allows for the evaluation of any existing CRS by treating it as a ``black box.''  That is, it does not require access to the source code or assume knowledge of its inner workings---it merely relies on observable behavior. 
Setting up an existing CRS with our simulator involves the following steps:

\begin{enumerate}[leftmargin=*]
    \item \textbf{Prepare domain and item collection}: A config file with domain-specific slot names must be prepared for the preference model.  Additionally, a file containing the item collection is required.
    \item \textbf{Provide preference data}: Preference data is consumed in the form of item ratings (user ID, item ID, and rating triples). 
	\item \textbf{Dialogue sample}: A small sample of dialogues with the CRS needs to be collected.  The sample size depends on the complexity of the system, in terms of action space and language variety, but is generally in the order of 5-50 dialogues.
	\item \textbf{Define interaction model}: A config file containing the space of user and agent intents (i.e., possible actions), as well as the set of expected agent responses for each user intent, is required for the interaction model. The baseline (CRSv1) interaction model shipped with the UserSimCRS library offers a  starting point, which may be further tailored according to the behavior and capabilities of the given CRS. 
	\item \textbf{Annotate sample}: The sample of dialogues must contain utter\-ance-level annotations in terms of intents and entities, as this is required to train the NLU and NLG components. Note that the slots used for annotation should be the same as the ones defined in the domain file (cf. Step 1) and intents should follow the ones defined in the interaction model (cf. Step 4.).
	\item \textbf{Define user model/population}: Simulation is seeded with a user population that needs to be characterized, for example, in terms of the different contexts (e.g., weekday vs. weekend, alone vs. group setting) and personas (e.g., patient and impatient users). Further, the number of users to be generated is to be specified. Each user will have their own preference model, which may be instantiated by grounding it in actual preferences (i.e., the ratings dataset given in Step 2). 
	\item \textbf{Train simulator}: The NLU, NLG, and response generation components of the simulator are trained using the annotated dialogue sample. 
	\item \textbf{Run simulation}: Running the simulator will generate a set of simulated conversations for each user with the CRS and save those to files. 
	\item \textbf{Perform evaluation}: Evaluation takes the set of simulated dialogues generated in the previous step as input, and measures the performance of the CRS in terms of the metrics implemented in DialogueKit.
\end{enumerate}

\begin{figure}
    \centering
    \includegraphics[width=0.45\textwidth]{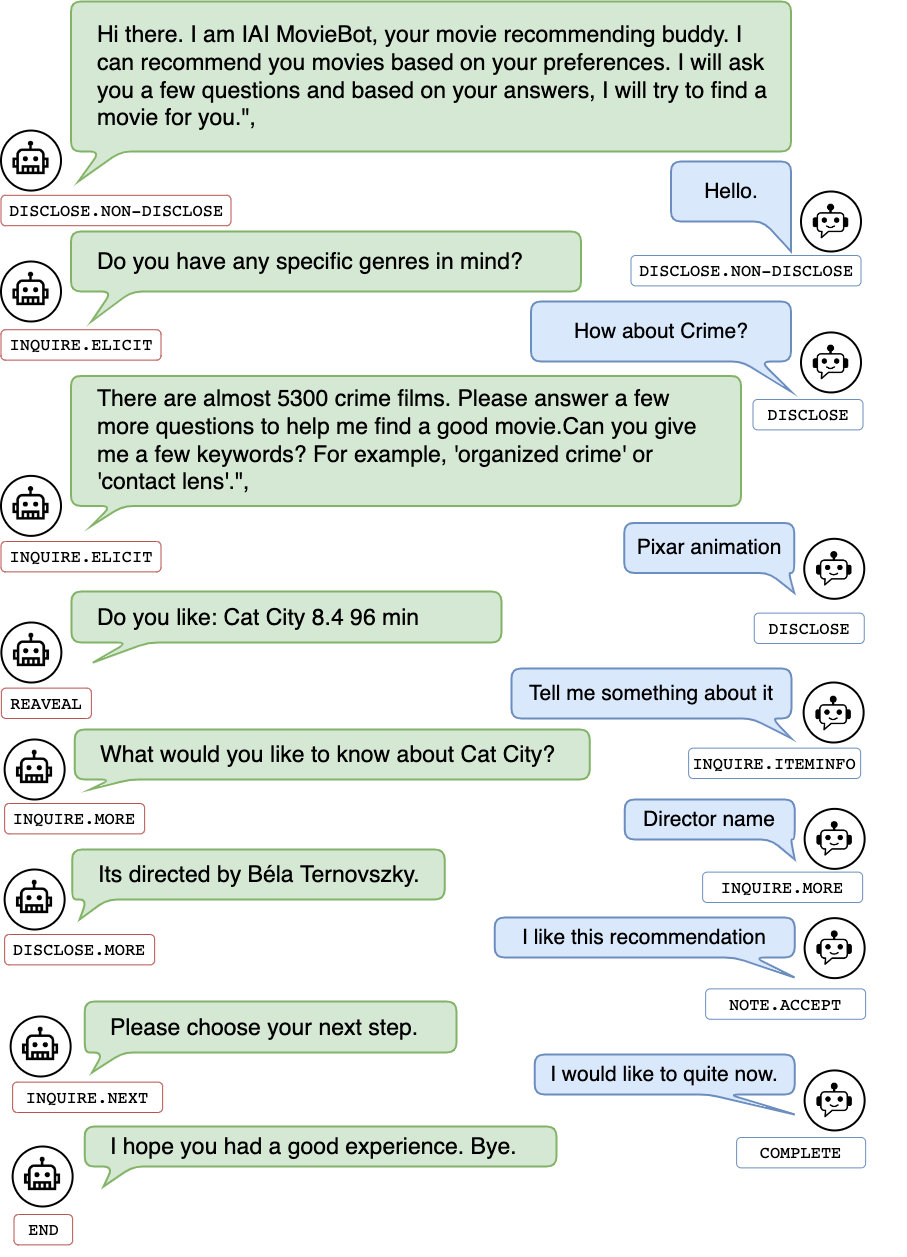}
    \vspace*{-\baselineskip}
    \caption{Sample dialogue between IAI MovieBot (Left, in green) and the user simulator (Right, in blue).}
    \label{fig:example_dialogue}
\end{figure}

\subsection{Case Study}
\label{sec:system:casestudy}

To see our user simulator in action, we conducted a case study with IAI MovieBot~\citep{Habib:2020:IMC},\footnote{\url{https://github.com/iai-group/MovieBot}} which is an open-source conversational movie recommender system.
This required creating a connector agent in DialogueKit, which can talk to IAI MovieBot via a RESTful API. 
We followed the steps listed above to prepare the user simulator.  This included collecting a sample of 8 dialogues, configuring the domain (with \emph{title}, \emph{genre}, and \emph{keyword} as slots), and annotating user and system utterances using intents (according to our CRSv1 interaction model) and slot-value pairs.
As it can be seen from the sample dialogue in Fig.~\ref{fig:example_dialogue}, the simulator could successfully complete dialogues with the CRS, mimicking the behavior of users observed in the training data it was exposed to.

\section{Conclusion and Future Directions}
\label{sec:concl}

We have presented a user simulation toolkit, organized into two Python libraries around general dialogue management and specific user simulation functionality, to facilitate research on both conversational recommender systems and simulation-based evaluation.  The toolkit is shipped with solid baseline models for each of the components, a detailed set of instructions, and a working example with an existing CRS, in order to make it easy for researchers and developers to start conducting simulation-based experiments.
Future work is concerned with extending the components with additional modeling options, including alternatives to agenda-based simulation.  We also plan to evaluate additional existing CRSs to ensure that our framework generalizes to diverse systems. \\

\noindent
\emph{\textbf{Acknowledgment.}}
We thank Nolwenn Bernard for her extensive contributions to the toolkit, made after the submission of this paper.

\bibliographystyle{ACM-Reference-Format}
\bibliography{wsdm2023-usersimcrs.bib}

\end{document}